# Single-frequency integrated laser on erbium-doped lithium niobate on insulator

Zeyu Xiao,[1] Kan Wu[1,*], Minglu Cai,[1] Tieying Li,[1] and Jianping Chen[1]

[1]State Key Laboratory of Advanced Optical Communication Systems and Networks, Shanghai Jiao Tong University, Shanghai 200240, China
*Corresponding author: kanwu@sjtu.edu.cn

**The erbium-doped Lithium niobate on insulator (Er:LNOI) platform has great promise in the application of telecommunication, microwave photonics, and quantum photonics due to its excellent electro-optic, piezo-electric, nonlinear nature as well as the gain characteristics in the telecommunication C-band. Here, we report a single-frequency Er:LNOI integrated laser based on dual-cavity structure. Facilitated by the Vernier effect and gain competition, the single-frequency laser can operate stably at 1531-nm wavelength with a 1484-nm pump laser. The output laser has a power of 0.31 μW, a linewidth of 1.2 MHz, and a side mode suppression ratio (SMSR) of 31 dB. Our work allows the direct integration of this laser source with existing LNOI components and paves the way for a fully integrated LNOI system.**

Lithium niobate on insulator (LNOI) or thin-film lithium niobate (TFLN) is a newly emerged integrated platform [1, 2]. As a material with exceptional electro-optic effect, piezo-electric effect, nonlinearity and wide transparency window, various important functions have been realized on LNOI platform including electro-optical modulators [3-5], acousto-optic devices [6, 7], Kerr microresonators [8-10], second harmonic generation [11, 12], and supercontinuum generation [13, 14].

As the lithium niobate (LN) material itself does not have gain properties, the integrated lasers and amplifiers are not supported in the pure LN platform. Many works have been focused on the erbium-doped thin-film LN materials, since the erbium ions can provide significant gain in the desired telecom C-band. In a recent research [15], the ion implantation method is performed on LNOI wafer. Limited by the low implantation concentration and non-uniform distribution of $Er^{3+}$, the fabricated micro-ring resonator does not exhibit laser emission. An alternative approach is to directly prepare Er-doped LN crystal, and then fabricate an erbium-doped LNOI (Er:LNOI) wafer by ion-cutting technology [16]. Thanks to the high doping concentration of $Er^{3+}$ ions and good mode confinement, small-signal gain of more than 5 dB/cm has been achieved in Er:LNOI waveguide amplifiers [16]. With same technology, microcavity-based lasers have been demonstrated in Er:LNOI wafer with an ultralow threshold power [17-20]. The excellent performance of Er:LNOI amplifiers and lasers clearly demonstrates the advantages of Er:LNOI platform over other erbium-doped integrated platforms.

For laser demonstrations on Er:LNOI platform, most works report multi-longitudinal-mode operation as a single ring or whispering gallery mode (WGM) resonator is adopted [17-20]. Very recently, Gao has shown a single-frequency lasing operation with two coupled WGM resonators on Er:LNOI. But the output power is still low, only ~50 nW [21]. A taper fiber is utilized to couple the pump and extract the lasing signal, which may also limit the application in a fully integrated system. Therefore, it is highly desired to achieve a single-frequency Er:LNOI laser with higher output power and a more integrated coupling scheme.

In this paper, we demonstrate a single-frequency integrated laser based on Er:LNOI platform. The dual-cavity structure is applied so that only one resonant mode can oscillate in the $Er^{3+}$ gain bandwidth through Vernier effect [22]. One of two micro-cavities has a very long cavity length so that more pump power can be absorbed and thus a higher laser power can be achieved. The micro-cavities and bus waveguide are all integrated on the same chip which enables a convenient and stable coupling of pump and signal. Experimentally, the laser can operate in a single-frequency state with a lasing wavelength near 1531 nm, an output power of 0.31 μW, a pump threshold of 13.5 mW, a linewidth of 1.2 MHz and a side mode suppression ratio (SMSR) of 31 dB. Our work paves the way of a fully integrated LNOI system with both active and passive components, which may find immediate applications in optical communications and signal processing.

The operation principle of Er:LNOI dual-cavity laser is shown in Fig. 1(a). The laser consists of a short micro-ring cavity (Cavity (a)) with a diameter of 200 μm, and a long cavity (Cavity (b)) with a length of ~1.2 cm. Cavity (a) is embedded in cavity (b) with two pulley coupling regions. The pulley coupling is designed so that cavity (a) works under over coupling state. Cavity (b) is designed to be much longer than cavity (a) so that more pump power can be stored in the cavity and the output power can be increased as well. Cavity (a) has a shorter length and a larger free spectral range (FSR, denoted as $F_a$ ~ 200 GHz) and cavity (b) has a longer length and a smaller FSR (denoted as $F_b$ ~ 10 GHz). Such a combination of different FSRs also enables a simple matching of the resonance peaks between two cavities. As a result, only the frequency which is located in the Er gain bandwidth and matches the resonance peaks of both cavities can oscillate, as illustrated in Fig. 1(b). This dual-cavity structure is inspired by the works of filter-driven four-wave mixing laser [23-25], which are used in the investigation of Kerr micro-combs.

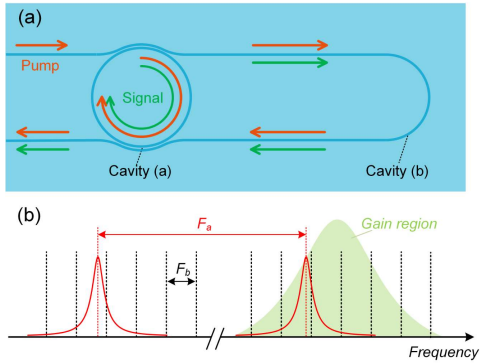

Fig.1. (a) Operation principle of dual-cavity Er:LNOI laser. (b) Vernier effect determines the oscillation frequency with two different FSRs.

The laser is fabricated on an $Er^{3+}$-doped LNOI wafer with 1 mol% doping concentration ( ~$1.89 \times 10^{20}$ $cm^{-3}$, Z-cut). The main fabrication process of Er:LNOI wafer is briefly described as follows. An erbium-doped LN wafer is first prepared with crystal growth process. Then it is bonded to a holder wafer with 2-μm silica layer and 400-μm silicon substrate by ion-cutting technology. The wafer is finally polished to obtain the desired thickness of LN layer. The laser device is fabricated by using drying etching technology with chromium (Cr) and hydrogen silsequioxane (HSQ) masks and inductively coupled plasma-reactive ion etching (ICP-RIE). The photograph of the fabricated chip is shown in Fig. 2(a). The scanning electron microscopic (SEM) image of the waveguide cross section is shown in Fig. 2(b). The ridge waveguide has a top width of 1.4 μm, an etching depth of 410 nm and a wedge angle of ~75°. The waveguide supports fundamental transverse-electric (TE) and transverse-magnetic (TM) modes. The SEM image of pulley coupling region is shown in Fig. 2(c). The coupling region has a coupling length of 96 um and a gap of 300 nm.

To evaluate the Q factor of the device with coupled dual-cavities, the transmission spectrum of the device is measured by a tunable continuous-wave laser. A polarization controller (PC) is used to adjust the polarization of the input laser and make the intracavity field transmit in TE mode. The tunable laser is scanned from 1528.5 nm to 1533.5 nm with a sweeping step of 0.1 pm. The output power of the laser is set to −10 dBm to avoid the resonance drift caused by photorefractive and thermo-optic effects [26, 27]. The measured transmission spectrum of the cold cavity is exhibited in Fig. 2(d). It is featured by equally spaced resonance peaks with large FSR of 200 GHz (cavity (a)) and small FSR of ~10 GHz (cavity (b)). Different from a single ring with bus waveguide, such a dual-cavity structure shows transmission peaks rather than dips in the resonance frequencies [28]. In Fig. 2(d), three resonance peaks with 200-GHz FSR are identified, marked by *Res*.1, *Res*.2 and *Res*.3 respectively. The zoomed transmission spectrum near *Res*. 2 near 1531.1 nm is presented in Fig. 2(e). Fitted by the Lorentz function, it has a linewidth of 3.9 GHz, corresponding to a loaded Q of $5\times10^4$.

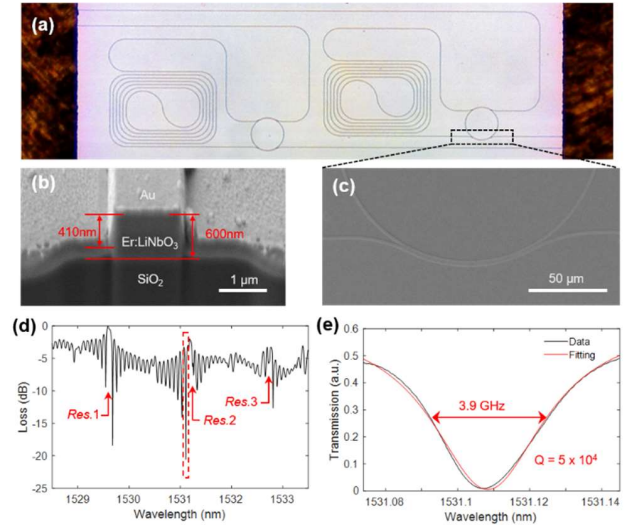

Fig.2. (a) Micrograph of the Er:LNOI laser with coupled dual-cavities. The SEM images of (b) the cross section of ridge waveguide, and (c) pulley coupling region. (d) Transmission spectrum of Er:LNOI dual-cavities. (e) A zoomed spectrum near *Res*. 2 near 1531 nm (black), and Lorentzian fit (red), exhibiting a *Q* factor of $5\times10^4$.

The experimental setup of Er:LNOI laser is shown in Fig. 3(a). A homemade pump laser centered at 1484 nm is chosen as it has better pumping efficiency than 980 nm due to the better mode overlapping [29]. The pump laser propagates through a polarization controller and is then coupled into the device via a lensed fiber. A TEC module is placed under the chip to stabilize the chip temperature. The output signal of the chip is collected by a second lensed fiber, and a wavelength-division multiplexer (WDM) is used to separate the residual pump light (1484 nm) and the signal light (1531 nm). The signal light is detected by the optical spectrum analyzer (OSA) with a resolution of 0.02 nm. In the experiment, the device shows strong green up-conversion fluorescence, as shown in the inset of Fig. 3(a). The luminescence of cavity (a) proves that the 1484-nm pump light is effectively coupled into the cascaded micro-ring cavity. Facilitated by the Vernier effect and gain competition, the $Er^{3+}$-doped LNOI micro-laser can operate stably in single-longitudinal-mode state. Fig. 3(b) shows the measured spectra in the range of 1480 nm to 1580 nm at pump power of 13.3 mW, 14.3 mW and 15.8mW respectively. Due to the limited isolation of the WDM, the leaked pump light of 1484 nm can still be detected by OSA. Single-frequency laser emission is observed near 1531 nm, which is consistent with the wavelength of *Res*.2 in Fig. 2(d). The gain peak of the erbium ion near 1550 nm also forms a small spontaneous emission envelope on the spectrum, but no laser excitation exists. The zoomed spectrum of signal light at pump power of 15.8mW is shown in Fig. 3(c), exhibiting an SMSR of 31dB and an output power of 0.31 μW. Currently, it is the highest SMSR and output power achieved by single-frequency integrated laser on Er:LNOI platform.

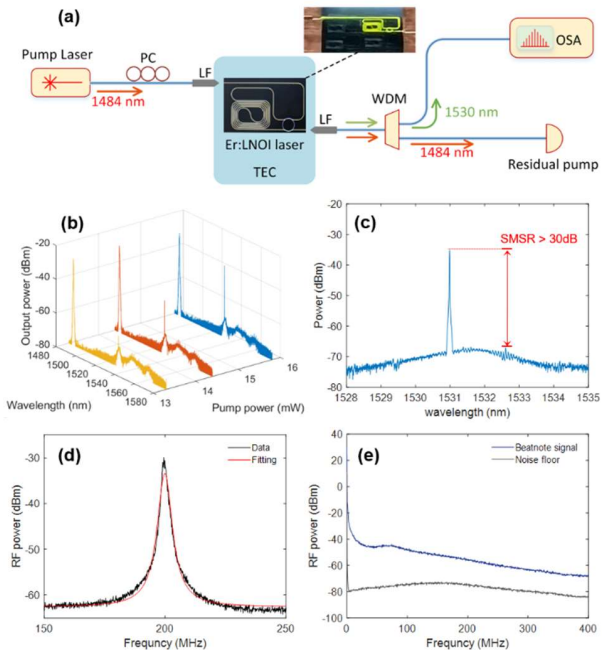

Fig.3. (a) Experimental setup. Inset: The photograph of operating laser, clearly indicating the green up-conversion fluorescence. PC: polarization controller; LF: lensed fiber; WDM: wavelength-division multiplexing; OSC: oscilloscope. (b) Emission spectra of Er:LNOI laser with a pump power of 13.3 mW, 14.3 mW and 15.8mW respectively. (c) Zoomed spectrum of the signal light, exhibiting a SMSR > 30 dB. (d) The measured self-heterodyne signal (black) of the Er:LNOI laser, and Lorentz fit (red). (e) The measured beat-note signal of laser emission (blue), and the noise floor of the measure system (grey).

The linewidth of laser is measured by self-heterodyne measurement [30]. The laser output is amplified by an erbium-doped fiber amplifier (EDFA), and split into two light paths with different delays for self-heterodyne measurement. An acousto-optic modulator is used in one branch to provide a frequency shift of 200 MHz, which avoids the interference from low-frequency band. The measurement result of the self-heterodyne signal is shown in Fig. 3(d), along with Lorentz fitting curve. It can be obtained the full-width-at-half-maximum (FWHM) linewidth of the demonstrated single-frequency laser is 1.2 MHz. To illustrate that our laser is operating in pure single-mode mode, rather than the co-existence of TE and TM modes, we directly measure the beat-note signal of the emission light, as shown in Fig. 3(e). No beat-note signal is observed in the 400 MHz range, indicating that only the fundamental TE mode is excited in the cascade micro-cavity.

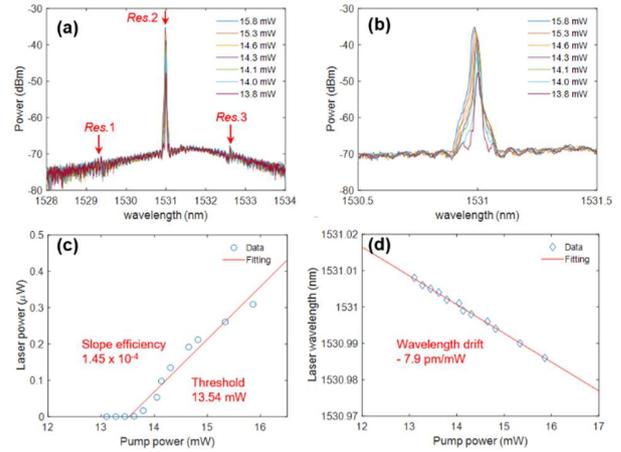

Fig.4. (a) The measure emission spectrum with different on-chip pump power. (b) The zoomed-in spectrum of the signal light. (c) The emission power and (d) wavelength drift of signal light versus different pump power, and fitted by linear function (red curve).

The laser performance under different pump power is then investigated as shown in Fig. 4. Fig. 4(a) shows the spectral evolution of the laser output at different on-chip pump power. In Fig. 4(a), the spectra have two small envelopes near 1529 nm and 1533 nm because the optical field of cavity (a) is coupled into cavity (b). The wavelengths of these two envelopes match well with the wavelengths of *Res*.1 and *Res*.3 in Fig. 2(d), and the laser emission wavelength is matched to the wavelength of *Res*.2 in Fig. 2(d). Compared to *Res*.1 and *Res*.3, *Res*.2 has a relatively low intracavity loss due to the Vernier effect and builds up a stable lasing signal in the gain competition. The laser output power under different pump power is summarized in Fig. 4(c). After linear fitting of the signal power, the power threshold and slope efficiency of the laser is 13.54 mW and $1.45\times10^{-4}$. Compared to previous works [17-20], our Er:LNOI micro-laser has a relatively high threshold. It is because the homemade 1484-nm laser has a wide spectrum of 0.5 nm and only little pump power is coupled to the cavity due to the resonance condition. If a narrow linewidth pump laser is used and pumped at the resonant frequency, the threshold of the laser can be reduced significantly.

Fig. 4(b) shows the zoomed view of the signal light spectra under different pump power. Interestingly, the lasing wavelength experiences a blue shift with the increase of pump power. Fig. 4(d) summarizes the wavelength variation versus the pump power, and the slope of linear wavelength drift is fitted to be −7.9 pm/mW. Obviously, the refractive index of the cavity changes with the increase of pump power and leads to a shift of the resonance frequency. There are three mechanisms that can change the refractive index of Er:LNOI waveguide, i.e., thermo-optic effect, photorefractive effect and carrier effect in erbium-doped waveguide. The carrier effect in erbium-doped fiber amplifier has a maximum pumped/unpumped refractive index change of $5.5\times10^{-8}$/dB of absorption at 1536 nm [31]. Therefore, the refractive index change caused by the carrier effect is negligible when increasing the pump power. The congruent single-crystalline LN exhibits a positive thermo-optic coefficient of $dn_e/dT \approx 3.8\times10^{-5}/K$ for the extraordinary polarization and $dn_o/dT \approx 0.69\times10^{-6}/K$ for ordinary polarization at room temperature and C-band wavelength range [32]. Since the on-chip laser is fabricated in a Z-cut Er:LNOI wafer, thermo-optic effect induced refractive index variation is mainly determined by $dn_o/dT$ due to TE mode transmission. The relatively small $dn_o/dT$ leads to a small red-shift of the lasing wavelength with increased pump power. The photorefractive effect decreases the refractive index with the increased pump power and thus leads to an increased FSR and a blue shift of the lasing wavelength. The photorefractive effect induced variation rate $\Delta n_o/n_o$ for ordinary polarization has an order of $\sim 1\times10^{-4}$ [33]. In the experiment, the laser wavelength has a blue drift with increased pump power, and the relative wavelength drift $\Delta\lambda/\lambda_o$ is about $1.4\times10^{-5}$, so it can be deduced that the wavelength drift is mainly caused by the photorefractive effect.

In summary, an on-chip single-frequency laser is demonstrated on Er:LNOI platform. The Vernier effect is introduced into the laser by a dual-cavity design. The single-frequency laser can consistently operate in 1531 nm with an output power of 0.31 μW, and a side mode suppression ratio (SMSR) of 31 dB. This work can provide an on-chip coherent light source on LNOI platform and can play an important role in a fully integrated LNOI system for the applications of optical communications and signal processing.

**Funding.** National Nature Science Foundation of China (NSFC) (No. 61922056, 61875122).

**Disclosures.** The authors declare no conflicts of interest.